\documentstyle[art12]{article}
\def \fr#1#2{\ifinner \innerfr{#1}{#2}  \else \outerfr{#1}{#2}\fi}
\def \innerfr#1#2{\ifmmode \ndmathfr{#1}{#2}
           \else \textfr{#1}{#2}\fi}
\def \outerfr#1#2{\ifmmode \dmathfr{#1}{#2}
           \else \textfr{#1}{#2}\fi}
\def \ndmathfr #1#2{ \raise 0.4 ex\hbox{${\scriptstyle {#1 \over #2}} $} }
\def \dmathfr  #1#2{ {#1 \over #2} }
\def \textfr   #1#2{\raise 0.4 ex\hbox{${\scriptstyle {#1 \over #2}}$}}
\def\fun#1#2{\lower3.6pt\vbox{\baselineskip0pt\lineskip.9pt
  \ialign{$\mathsurround=0pt#1\hfil##\hfil$\crcr#2\crcr\sim\crcr}}}
\def\centeron#1#2{{\setbox0=\hbox{#1}\setbox1=\hbox{#2}\ifdim
\wd1>\wd0\kern.5\wd1\kern-.5\wd0\fi
\copy0\kern-.5\wd0\kern-.5\wd1\copy1\ifdim\wd0>\wd1
\kern.5\wd0\kern-.5\wd1\fi}}
\def\centerover#1#2{\centeron{#1}{\setbox0=\hbox{#1}\setbox
1=\hbox{#2}\raise\ht0\hbox{\raise\dp1\hbox{\copy1}}}}
\def\centerunder#1#2{\centeron{#1}{\setbox0=\hbox{#1}\setbox
1=\hbox{#2}\lower\dp0\hbox{\lower\ht1\hbox{\copy1}}}}
\def \UNIT #1{{\ifmmode \UNITBODY {#1} \else $ \UNITBODY {#1} $\fi }}
\def \UNITBODY #1{\,{\rm #1}}

\def \GeV {\UNIT{GeV}}

\def \cm  {\UNIT{cm}}

\def \s   {\UNIT{s}}

\def \MSbarbasic {\overline{{\rm MS}}}
\def \MSbar {\ifmmode \MSbarbasic \else $\MSbarbasic$\fi }

\def \bra#1|{\mathopen{\langle#1\,|}}  
\def \braket#1|#2>{\langle#1\,|\,#2\rangle} 
\def \ket#1>{\mathclose{|\,#1\rangle}}  

\def \ltap{\;\centeron{\raise.35ex\hbox{$<$}}{\lower.65ex\hbox{$\sim$}}\;}
\def \gtap{\;\centeron{\raise.35ex\hbox{$>$}}{\lower.65ex\hbox{$\sim$}}\;}

\def \e{{\rm e}}
\def \epem{{\ifmmode \epembody \else $ \epembody $\fi }}
\def \epembody{ \e^+\e^- }

\def \rfj   #1#2#3#4{\jour {#1}\rf {#2}{#3}{#4}}

\relax
\newif\ifpreprint
\preprinttrue

\def\SECTIONHEAD#1{\vglue 0.6cm
{\elevenbf\noindent #1\hfil}
\vglue 0.4cm{}}

\def\SECTION#1{\refstepcounter{section}\SECTIONHEAD{\thesection. #1}}

\renewcommand{\it}{\elevenit}
\newcommand{\mainrm}{\elevenrm}

\font\tenbf=cmbx10
\font\tenrm=cmr10
\font\tenit=cmti10
\font\elevenbf=cmbx10 scaled\magstep 1
\font\elevenrm=cmr10 scaled\magstep 1
\font\elevenit=cmti10 scaled\magstep 1

\ifpreprint\else\fi

\renewenvironment{thebibliography}[1]
 { \elevenrm
   \begin{list}{\arabic{enumi}.}
    {\usecounter{enumi} \setlength{\parsep}{0pt}
     \setlength{\itemsep}{3pt} \settowidth{\labelwidth}{#1.}
     \sloppy
    }}{\end{list}}
\def\CITE#1{${}^{{\hbox{\cite{#1}}}}$}
\def\rfj#1#2#3#4{{\elevenit #1}\ {\elevenbf #2} (19#4) #3}
\def\title#1{{\elevenit #1}}

\begin{document}

\ifpreprint
   \begin{flushright}
      PSU/TH/118
   \end{flushright}
\fi
\begin{center}{{\tenbf SPIN IN PERTURBATIVE QCD;\\[3pt]
               COHERENT HARD DIFFRACTION\ifpreprint\footnote{Presented
                  at annual meeting of Division of Particles
                  and Fields of American Physical Society,
                  Fermilab 10--14 November 1992.}
               \fi
               \\}
\vglue 1.0cm
{\tenrm JOHN C. COLLINS\\}
\baselineskip=13pt
{\tenit Physics Department, Penn State University\\}
\baselineskip=12pt
{\tenit University Park PA 16802, U.S.A.\\}
\vglue 0.8cm
{\tenrm ABSTRACT}}
\end{center}
\vglue 0.3cm
{\rightskip=3pc
 \leftskip=3pc
 \tenrm\baselineskip=12pt
 \noindent
    I briefly review: (a) some recent developments in the theory of
    hard scattering in QCD with polarized beams, and (b) coherent
    hard diffraction (that is, hard scattering in diffractive
    events, with the Pomeron behaving in an apparently point-like
    fashion).
}

\baselineskip=14pt
\mainrm

\SECTION{Polarized Hard Scattering}

Interest in polarized hard scattering has increased
recently.  An important reason is that it appears feasible to
use polarized protons in the Relativistic Heavy Ion Collider at
Brookhaven\CITE{PW}.  The aim is an energy of 200+200 GeV at
a luminosity of $2\times 10^{32}\cm^{-2}\s^{-1}$ and a polarization of 70\%
for each beam.  So the theoretical study of polarization in
perturbative QCD is no longer academic. Polarized hard scattering
probes the spin-dependence, and hence the chiral properties, of
both the proton wave function and quark fragmentation, as is
particularly evident in recent work.

Rather intriguing are the measurements of spin asymmetries in
pion production at large $x_{T}$ in collisions of transversely
polarized protons with unpolarized hadrons. Leading twist QCD
predicts no asymmetry, so the large measured
asymmetries\CITE{pion} indicate large higher twist effects.

\SECTION{Twist 3}

Qiu and Sterman\CITE{QS} have recently shown that the first
non-leading twist contributions in hadron-hadron scattering (to
make jets or high $p_{T}$ photons etc) can be fitted into the
factorization framework.  They and other investigators, like
Jaffe and Ji\CITE{JJ}, have looked at the various contributions.

One particularly interesting contribution involves a quark-gluon
correlation function for the polarized proton.  Qiu and Sterman
argue that a pole in the hard scattering picks out a {\it
derivative} of the correlation function, and thereby enhances the
contribution at large parton $x$.  The flavor systematics appear
appropriate to the pion production results.

In general, it is rather difficult to extract the quark-gluon
correlation function in a model-independent way, since there are
two longitudinal momentum fractions involved on the polarized
side, and the kinematics of the hard scattering final state can
only determine the sum of these momentum fractions.

\SECTION{Spin-dependence of Fragmentation}

We now return to leading twist.  To treat deep-inelastic
scattering, for example, with a polarized proton, one must
equip the parton entering the hard scattering with a
helicity density matrix.

Transverse spin is particularly interesting because it
corresponds to off-diagonal terms in the helicity density matrix.
Unfortunately, the usual hard scattering coefficient is diagonal in
helicity, because of helicity conservation for massless quarks,
which is true to all orders of perturbation theory.  Hence
single transverse spin asymmetries are zero in leading twist for
inclusive deep inelastic scattering, etc\CITE{KPR}.

We can evade this result by performing a spin-sensitive
measurement on the outgoing quark jet. Nachtmann\CITE{Nacht} was
the first to suggest measuring the spin of a jet, for the
longitudinally polarized jets produced in neutrino
scattering. His idea
was recently rediscovered by Efremov, Mankiewicz and
Tornqvist\CITE{EMT}.
They define a `handedness' by $H \propto
\epsilon _{ijk} k_{1}^{i}k_{2}^{j}k_{3}^{k}.$ Here $k_{1}$,
$k_{2}$ and $k_{3}$ are 3-momenta of, for example, two leading
particles and the jet itself, as measured in the center-of-mass
of the hard scattering. The average
handedness is equal to the helicity
$\lambda $ of the parton initiating the jet times a nonperturbative
analyzing power. The analyzing power must be measured, of course,
but it is universal between different processes.

One can apply the same idea to jets initiated by transversely
polarized quarks, as shown by Efremov et al.\CITE{EMT} and by
myself and collaborators\CITE{trfrag}.  There, we consider
$\epsilon _{ijk} s^{i}k_{1}^{j}k_{2}^{k}$, which is a scalar.  Here $s$ is the
transverse spin of the quark. Notice that it is now only
necessary to measure two outgoing momenta: two leading
hadrons of opposite charges, or one leading hadron and the
jet momentum. This is in distinct contrast to the decay of a
real particle, where it would be necessary to do a three
body measurement.

This quantity then allows a $\cos\phi $ dependence on the azimuthal
angle between the transverse spin of the quark and normal to the
plane of two leading particles.  Again there is a nonperturbative
analyzing power to be measured.  Many possibilities now appear.
There should be twist 2 asymmetries in deep inelastic scattering
for an unpolarized electron on a transversely polarized proton,
with measurement of the current quark jet, and in jet production
in singly polarized proton-proton scattering. There can also a
jet-jet correlation in $e^{+}e^{-}$ annihilation at LEP.

A simple model for nonperturbative QCD is the Georgi-Manohar
sigma model of quarks and pions\CITE{GM}.  Ladinsky and
I\CITE{fragcalc} have done a calculation of the transverse
spin asymmetry of quark fragmentation in this model.  We
find a large asymmetry that relies on chiral symmetry
breaking and large imaginary parts in nonperturbative strong
interactions.

\SECTION{Coherent Hard Diffraction}

I now describe recent work with Frankfurt and
Strikman\CITE{FS,CFS} on what we call `coherent hard
diffraction'.  Defining the term will explain what we are doing.

Diffraction is a process like $p+p\to p+X$ where one of the
initial hadrons is almost undeflected and retains a fraction
$1-\xi $
of its initial momentum, with $\xi $ close to zero.  The
mass-squared of the $X$ system is $s'=\xi s$.  Diffraction is
generally attributed to Pomeron exchange between the diffracted
proton and $X$; effectively, $X$ is the result of a
Pomeron-proton collision. UA8\CITE{UA8} has data at the CERN
collider with $\sqrt {s'}$ in the 130 to 200 GeV range, and $\sqrt s=630
\GeV$. They then look for jets in $X$, which gives an
example of `hard diffraction'.

The simplest model for hard diffraction is due to Ingelman and
Schlein\CITE{IS}.  They assume that there are parton
distributions in the Pomeron, and apply the usual hard scattering
formalism.  With a reasonable ansatz for the gluon distribution,
one gets an appropriate size to fit the UA8 data, and so UA8 make
at least a rough measurement\CITE{UA8} of the gluon
distribution in a Pomeron.

However, there is no proof of the Ingelman-Schlein model. Indeed,
Frankfurt and Strikman\CITE{FS} made a calculation of hard
diffraction with two gluon exchange as a model for the Pomeron.
They found a {\it leading twist}  contribution where all the
momentum of the Pomeron goes into the jets.  Experimentally, this
looks as if there is a $\delta (x-1)$ term in the gluon distribution in
the Pomeron.  This was predicted in advance of the UA8 data, and
the data, after allowing for smearing by gluon radiation and by
the detector, etc, shows strong evidence for such a delta
function, in addition to the Ingelman-Schlein term.

\sloppypar
This process we call `coherent hard diffraction'. As I have
shown, in collaboration\CITE{CFS} with Frankfurt and Strikman, it
results from a breakdown of the factorization theorem.  The proof
of the factorization theorem requires an inclusive sum over the
spectators to the hard scattering, but the diffractive condition
on the final state prohibits the inclusive sum.

One striking part of the breakdown of factorization is that the
delta function term is not universal.  As shown by a
calculation by Donnachie and Landshoff\CITE{DL}, coherent hard
diffraction in deep inelastic lepton-hadron scattering is higher
twist.

\SECTION{Acknowledgments}

This work was supported in part by the U.S. Department of Energy
under grant DE-FG02-90ER-40577, and by the Texas National
Laboratory Research Commission.  I would like to thank many
colleagues for discussions and collaborations.

{\bf Note:}
Twist-3 parton densities were earlier considered by
Shuryak and Vainshtein\cite{Shuryak}.  Another way of
probing the polarization of a jet is from the decays of $\Lambda $s
in the jet\cite{AM,CPR}.


\SECTION{References}
%
%


\begin{thebibliography}{99}

\bibitem{PW} G. Bunce et al., \rfj{Particle World}{3}{1}{92}.

\bibitem{pion} E704 Collaboration: D.L. Adams et al.,
   \rfj{Phys.\ Lett.}{B264}{462}{91},
   \rfj{Phys.\ Lett.}{B261}{201}{91},
   \rfj{Phys.\ Lett.}{B276}{531}{92}, and
   references quoted there to earlier experiments at lower
   energies.

\bibitem{QS} J.-W. Qiu and G. Sterman, \rfj{Nucl.\
   Phys.}{B378}{52}{92}.

\bibitem{JJ} X.-D. Ji, \rfj{Phys.\ Lett.}{B289}{137}{92};
   I. Balitsky and V. Braun, \rfj{Nucl.\ Phys.}{B361}{93}{91}.

\bibitem{KPR} E.g., G. Kane, J. Pumplin, and W. Repko,
   \rfj{Phys.\ Rev.\ Lett.}{41}{1689}{78}.

\bibitem{Nacht} O. Nachtmann, \rfj{Nucl.\ Phys.}{B127}{314}{77}.

\bibitem{EMT} A.V. Efremov, L. Mankiewicz and N.A.
   T\"ornqvist, \rfj{Phys.\ Lett.}{B284}{394}{92}.

\bibitem{trfrag} J.C. Collins, \title{Fragmentation of
   Transversely Polarized Quarks Probed in Transverse Momentum
   Distributions}, Penn State preprint PSU/TH/102; R. Carlitz, J.C.
   Collins, S. Heppelmann, R. Jaffe, X. Ji, G. Ladinsky,
   \title{Measuring
   Transversity Densities in Singly Polarized Hadron-Hadron
   Collisions}, Penn State preprint PSU/TH/101, in preparation.

\bibitem{GM} A.\ Manohar and H.\ Georgi \rfj{Nucl.\
   Phys.}{B234}{189}{84}.

\bibitem{fragcalc} J.C. Collins and G.A. Ladinsky, \title{On
   $\pi $-$\pi $ Correlations in Polarized Quark Fragmentation
   Using the Linear Sigma Model}, Penn State preprint
   PSU/TH/114 in preparation.

\bibitem{FS} L.I. Frankfurt and M. Strikman, \rfj{Phys.\ Rev.\
   Lett.}{64}{1914}{89}; L.I. Frankfurt, \title{Hard Diffraction
   Processes at Colliders}, talk at F.A.D. Meeting at Dallas, TX,
   14 March 1992.

\bibitem{CFS} J.C. Collins, L.I. Frankfurt and M. Strikman,
   \title{Diffractive Hard Scattering with a Coherent Pomeron}, Penn
   State preprint PUS/TH/116.

\bibitem{UA8} A. Brandt et al., [UA8 Collaboration],
   \title{Evidence for a Super-Hard Pomeron Structure},
   submitted to Phys.\ Lett.\ 1992.

\bibitem{IS} G. Ingelman and P. Schlein,
   \rfj{Phys.\ Lett.}{152B}{256}{85}.

\bibitem{DL} A. Donnachie and P.V. Landshoff,
   \rfj{Phys.\ Lett.}{285B}{172}{92}.

\bibitem{Shuryak} E. Shuryak and Vainshtein, \rfj{Nucl.\
Phys.}{201}{141}{82}.

\bibitem{AM} X. Artru and M. Mekhfi,
   \rfj{Z. Phys.\ C}{45}{669}{90}.

\bibitem{CPR} J.L. Cortes, B. Pire and J.P. Ralston,
   \rfj{Z. Phys.\ C}{55}{409}{92}.

\end{thebibliography}
\end{document}